\nonstopmode
%
%
%
%
%
%

\def\twelvepoint{\magnification=\magstep1}

%
\twelvepoint
%
%

\def\imke11space{\baselineskip=11pt}
%
%
%
%
\vsize=22.7 true cm
\hsize=16.7 true cm
\voffset=0.0 true cm
\hoffset=0.0 true cm
\tolerance=100000
\parskip=0.2cm
\parindent=20pt
\raggedbottom
\def\onesided{\hoffset=0.0 true cm}
\def\twosided{\output={
   \ifodd\pageno \hoffset=0.0true cm \else \hoffset=0.0true cm \fi
   \shipout\vbox{
      \makeheadline
      \pagebody
      \makefootline
   }
   \advancepageno
   \ifnum\outputpenalty>-20000 \else\dosupereject \fi
}}
%
\onesided
%
%

%
%
\def\pp{\noindent \parshape 2 0truein 6.2truein 0.5truein 5.3truein}
%
%
%
\def\ref#1{{\par\parindent=0pt\pp #1\par}}
\def\apjref#1;#2;#3;#4 {{\parindent=0pt\par\pp#1, {\it #2}, {\bf #3}, #4. \par}}
%
%

%
%

%
%
\def\leaderfill{\leaders\hbox to 1em{\hss.\hss}\hfill}
%
%
\def\ltsima{$\; \buildrel < \over \sim \;$}
\def\simlt{\lower.5ex\hbox{\ltsima}}
%
%
\def\gtsima{$\; \buildrel > \over \sim \;$}
\def\simgt{\lower.5ex\hbox{\gtsima}}
%
%
\def\square{{\vbox{\hrule height.4pt
        \hbox{\vrule width.4pt height6pt \kern6pt
           \vrule width.4pt}
        \hrule height.4pt}}}
%
%

%
%

\def\deg{$^\circ$}

\def\et{{\it et al. }}

\def\ms{$\mu$m }

\def\list#1#2{\hangafter=1 \hangindent=1.75pc}
%
%
\def\chaphead{}
\def\badbreak{\penalty500}
\def\increment#1{\global\advance#1 by 1}
\newcount\eqnumber
\eqnumber=1
\def\new{{\rm(\chaphead\the\eqnumber}\increment\eqnumber}
\def\eqnam#1#2{\badbreak\immediate\write1{Equation #2 {\chaphead\the\eqnumber}}
	\xdef#1{\chaphead\the\eqnumber}}
%
%
\newcount\fignumber
\fignumber=1
\def\fignam#1#2{\immediate\write1{Figure #2 {\the\fignumber}}
	\xdef#1{\the\fignumber}\increment\fignumber}
%
%
\newcount\tablnumber
\tablnumber=1
\def\tablnam#1#2{\immediate\write1{Table #2 {\the\tablnumber}}
	\xdef#1{\uppercase\expandafter{\romannumeral\the\tablnumber}}
	\increment\tablnumber}
%
%
\catcode`@=11
\def\footnote#1{\edef\@sf{\spacefactor\the\spacefactor}#1\@sf
      \insert\footins\bgroup\eightpoint
      \interlinepenalty100 \let\par=\endgraf
        \leftskip=0pt \rightskip=0pt \baselineskip=10pt
        \splittopskip=10pt plus 1pt minus 1pt \floatingpenalty=20000
        \smallskip\item{#1}\bgroup\strut\aftergroup\@foot\let\next}
\skip\footins=12pt plus 2pt minus 4pt 
\dimen\footins=30pc 
\newcount\notenumber
\def\clearnotenumber{\notenumber=0}

\def\note{\advance\notenumber by 1 \footnote{$^{[ \the\notenumber ]}$}}

\clearnotenumber
%
%

\font\eightrm=cmr8
\font\sixrm=cmr6

\font\ninei=cmmi9
\font\eighti=cmmi8
\font\sixi=cmmi6
\skewchar\ninei='177 \skewchar\eighti='177 \skewchar\sixi='177

\font\ninesy=cmsy9
\font\eightsy=cmsy8
\font\sixsy=cmsy6
\skewchar\ninesy='60 \skewchar\eightsy='60 \skewchar\sixsy='60

\font\eightbf=cmbx8
\font\sixbf=cmbx6



\font\eightsl=cmsl8

\font\eightit=cmti8

\font\eightsl=cmsl8

\def\eightpoint{\def\rm{\fam0\eightrm}%
  \textfont0=\eightrm \scriptfont0=\sixrm \scriptscriptfont0=\fiverm
  \textfont1=\eighti \scriptfont1=\sixi \scriptscriptfont1=\fivei
  \textfont2=\eightsy \scriptfont2=\sixsy \scriptscriptfont2=\fivesy
  \textfont3=\tenex \scriptfont3=\tenex \scriptscriptfont3=\tenex
  \def\it{\fam\itfam\eightit}%
  \textfont\itfam=\eightit
  \def\sl{\fam\slfam\eightsl}%
  \textfont\slfam=\eightsl
  \def\bf{\fam\bffam\eightbf}%
  \textfont\bffam=\eightbf \scriptfont\bffam=\sixbf
   \scriptscriptfont\bffam=\fivebf
  \normalbaselineskip=9pt
  \let\sc=\sixrm
  \let\big=\eightbig
  \normalbaselines\rm}
\input epsf.tex
\parindent=0pt

\centerline {\bf Longitude-Resolved Imaging of Jupiter at $\lambda=2$ cm}
\medskip
\centerline{R.J. Sault, Chermelle Engel, and Imke de Pater}
\medskip
\centerline {Australia Telescope National Facility}
\centerline {CSIRO, Epping, NSW 2121, Australia}
\medskip
\centerline {Astronomy Department, 601 Campbell Hall}
\centerline {University of California, Berkeley CA 94720}
\vfill
Pages: 28

Tables: 0

Figures: 5

{\bf Proposed Running Head:} Longitude-Resolved Imaging of Jupiter
\bigskip
{\bf Editorial correspondence to:}

R.J. Sault\hfill\break
Australia Telescope National Facility\hfill\break
Locked Bag 194\hfill\break
NARRABRI, NSW, 2390\hfill\break
AUSTRALIA
Email: rsault@atnf.csiro.au
\eject

{\bf Abstract} 

We present a technique for creating a longitude-resolved image of Jupiter's
thermal radio emission. The technique has been applied to VLA data taken on
25 January 1996 at a wavelength of 2 cm.  A comparison with infrared data
shows a good correlation between radio hot spots and the 5 \ms hot spots
seen on IRTF images. The brightest spot on the radio image is most likely
the hot spot through which the Galileo probe entered Jupiter's atmosphere.
We derived the ammonia abundance ($=$ volume mixing ratio) in the hot spot,
which is $\sim 3 \times 10^{-5}$, about half that seen in
longitude-averaged images of the NEB, or less than 1/3 of the
longitude-averaged ammonia abundance in the EZ. This low ammonia abundance
probably extends down to at least the 4 bar level.

\bigbreak
{\bf 1. Introduction}

{\it Radio Astronomy}

Conventional radio interferometry images are integrated over many hours,
with 12 or more hours not being unusual. This is required both to meet the
required sensitivity and to use Earth rotation synthesis to achieve good sampling of
the Fourier plane. Consequently, imaging planets in the conventional way
rotationally smears any longitudinal structure.  In principle one can merge
together snapshots of the same rotational aspect of the planet, from
observations taken on different days.  This approach was probably first
used by de Pater (1980) when observing Jupiter. In practice, the
longitudinal smearing is still limited by the fineness of the rotational
phase bins that the data are broken into.  At decimetric wavelengths, the
spatial resolution of the data used to be a significant fraction of a
planetary radius, and extra smearing caused by rotation was not so severe.
These are typically tens of degrees of rotation.  To image the thermal
radiation of the planet itself, one requires a higher resolution, typically
1-2$''$ for Jupiter, or an integration time of less than 10 minutes. Such
an image has such poor signal-to-noise that even the North Equatorial Belt
(NEB) on Jupiter may not be recognized. To date, all radio wavelength
images of Jupiter's thermal emission have therefore been averaged over
longitude (i.e., integrated over hours rather than minutes of time).

Sault \et (1997) developed an alternative technique
for Jupiter's synchrotron emission, based on a
three-dimensional tomographic approach. This
technique cannot be used for the thermal emission from the disk of a
planet. 
The present paper describes a technique to form a radio ``map'' of a planet's
thermal radio emission which
avoids rotational smearing, but which still allows good sensitivity and
Fourier plane
sampling. The main motivation to develop
this technique has been the need to form radio maps of Jupiter.

\medskip

{\it Jupiter Science}

Infrared observations of Jupiter at 5 $\mu$m show that the NEB is not
a smooth belt, but rather contains numerous ``hot spots'' -- regions
which are much hotter than their surroundings. In these hot spots, one probes
relatively deep, hot, levels in Jupiter's atmosphere
because the infrared
opacity is relatively low here. A main source of opacity at these wavelengths is cloud
layers; i.e., the absence of cloud particles in hot spots lets one
probe much deeper layers. The general consensus therefore has been
that hot spots are dry regions on Jupiter, perhaps areas of
downdrafts: rising air on the planet cools off, and once the
temperature drops below the condensation temperature of ammonia gas,
droplets and clouds form. Above the clouds the air is dry, and if
there are places of downdrafts the air may indeed be dry and hence
devoid of cloud particles.

On 7 December 1995 at 22:10 UT the Galileo probe entered Jupiter's
atmosphere, by chance entering in a hot spot (at a jovicentric
latitude of 6.5$^\circ$ N and at $\sim 4.5^\circ$ West longitude
(Syst. III); e.g., Orton \et 1998).  During its descent, the probe
measured the atmospheric structure (temperature, pressure, density),
gas composition and cloud properties down to a depth of $\sim$ 20 bar.
Folkner \et (1998) derived the ammonia abundance profile in the hot
spot from the gradual and progressive attenuation of the probe signal
(at 1.4 GHz) while the probe was descending in Jupiter's
atmosphere. They determined an NH$_3$ abundance $\sim 3.6 \pm 0.5
\times$ the solar N value
\note{We use the solar elemental ratios for N,
O, and S given in 
Anders and Grevesse (1989) and for C given in 
Grevesse \et (1991).  The solar N/H value is $1.12 \times
10^{-4}$. The volume mixing ratio or mole fraction of solar N in Jupiter's
atmosphere is $1.97 \times 10^{-4}$. We use the word abundance as having
the same meaning as volume mixing ratio or mole fraction.
}
at pressures between 8 and 12 bar,
with a possible decrease at higher altitudes.  The NH$_3$ abundance
has since been obtained from the Galileo probe data (Mahaffy \et
1999), and was found to be $3.2 \pm 1.4 \times$ solar N, i.e.,
consistent with the results of Folkner \et (1998). Sromovsky \et (1998) deduced
an NH$_3$ profile at $P \simlt 3$ bar from data taken with the Net
Flux Radiometer (NFR) on board the Galileo probe. They found that NH$_3$ in
the hot spot was decreasing linearly from $0.5 \times$ solar N at 2
bar to $0.01 \times$ solar at $P \approx 0.5$ bar. The Galileo probe
further measured very low abundances of H$_2$S at $P
\simlt 10$ bar and H$_2$O at $P \simlt 20$ bar. Various researchers
have investigated downdrafts and planetary waves to explain the
altitude profiles of the condensible gases as measured by the Galileo
probe (see, e.g., Atreya \et 1996; Showman and Ingersoll 1998; Wong \et
1998; Showman and Dowling 2000; Friedson and Orton 1999).

De Pater \et (2001) (henceforth referred to as dP2001) observed
Jupiter at radio wavelengths (at 2, 3.6 and 6 cm wavelength) near the
time of the probe entry. Since the main source of opacity at radio
wavelengths is ammonia gas, dP2001 derived the disk-averaged
ammonia abundance profile from the data and compared this with the
Galileo probe results.  They could only reconcile an NH$_3$ abundance
$\sim 3.6 \pm 0.5 \times$ the solar N value at $P>8$ bar if the NH$_3$
abundance decreases globally at $P<4$ bar, and to subsolar values at
pressures $P \simlt 2$ bar. They further show (based upon
disk-resolved but longitude-averaged images) that the NH$_3$
abundance in the NEB is $\sim$ 50 -- 70\% of the value in the EZ
(Equatorial Zone), while it is subsolar in both bands at $P<2$ bar.
In the NEB this low abundance has to extend down to $P \sim$ 4 -- 6
bar.

With longitude-resolved radio images one could correlate the
occurrence of infrared hot spots with radio hot spots, if they were to
exist. Moreover, the main source of opacity at the two wavelength
ranges is different: at radio wavelengths one is sensitive to the
condensible gas, ammonia, while at IR wavelengths one is sensitive to
cloud particles.
So microwave observations would
provide information on the condensible gas, a necessary piece of
information for the development of dynamical models of Jupiter's
atmosphere.

To investigate the occurrence of hot spots, and in particular the hot spot
in which the Galileo probe descended, we developed an algorithm to
construct a longitude-latitude map of Jupiter, and applied this algorithm
to 2-cm data taken at the VLA near the time the Galileo probe descended into
Jupiter's atmosphere; the longitude-averaged maps were presented in dP2001. 
In this paper we discuss the algorithm (Section 2) and present the
observations (Section 3) and results (Section 4).

\bigbreak
{\bf 2. The mapping technique}

To understand the basis of our mapping technique, consider Fig.~1.
Figure 1a shows four identical crosses as they would appear on the
face of the planet. The viewing geometry distorts the appearance of
the crosses -- Fig 1b presents the crosses side-by-side to help
accentuate the distortions. On a sufficiently small scale, this
distortion can be approximated as a linear transformation of the
coordinate system.  That is, to first approximation, by appropriate
rotation and skew (length scaling), we can convert each of the four crosses
into the others. In particular, the appropriate transformation can
convert the three right-most crosses to that which would be seen face
on.

Applying this to the problem of imaging a rotating planet, all the
data (from different viewing aspects) of a sufficiently small facet
can be transformed such as they would appear as seen face on. All the
data can then be co-added. This assumes that there are no other
changes in the emission other than the viewing geometry, e.g.  local
solar time effects or limb effects are negligible.

In interferometry, where the basic data are samples in the Fourier domain,
these manipulations are conveniently done in the Fourier domain.
The manipulated visibility data can then be Fourier transformed to form
an image of the facet. The details of the manipulations that must be done
on the visibilities are described in Appendix A. Briefly, however, they
amount to three operations. In imaging a particular facet, the operations
that need to be applied to each visibility datum are:

\item{1} Apply a phase term to shift the data so that the phase center
is the center of the facet. This shift can account for the motion of the
facet across the face of the planet.

\item{2} Apply a linear transformation to the $u-v$ coordinates of the
data. This corrects for the viewing geometry of the facet as it
rotates across the planet. This transformation can readily cope with
an oblate spheroid (not just a sphere).  A correction for differential
rotation can also be made in this step: differential rotation can be
modelled as a latitudinal and time dependent stretching in the
longitudinal direction which is relevant for the gas giants. The winds
on Jupiter are largest near the equator, $\sim 100$ m/s; since
features here move $\simlt 3^\circ$ in 10 hours ($\simlt
1.3^{\prime\prime}$ as seen from Earth), we ignored differential rotation.

\item{3} Scale up the data (and consequently its noise variance) to account for
the change in the projected area. As the facet approaches the limb of
the planet, this scale factor approaches infinity, and the signal to
noise ratio of the data approaches zero.

Note that the details of each step changes as a function of time
(or viewing geometry). The steps can also readily handle changes in the 
observer-planet
distance and geometry during the course of the observation.

When imaging this facet, we weight the visibility data
inversely with the noise variance. This gives little weight to data
measured near the limb of the planet. Consequently the variance of the
resultant image is not particularly degraded by limb data. We further
use weighting schemes such as Briggs' robust weighting (1995)
to protect against excessively noisy data.

Before visibility manipulation and imaging, it is sensible to subtract
a uniform model of the planet, so that we are imaging only deviations
from this model. For example, in the data presented below, we have
subtracted a model corresponding to a constant temperature oblate
spheroid with limb darkening (see Butler and Bastian 1999 for the
corresponding visibility function), the best-fit parameters for which
were derived from the data. 

Mapping a large part of the planet (not a single small facet) can
be achieved by stitching together a number of facets.  The stitching proceeds by
regridding the facets to a common mapping geometry (e.g. Mercator) and
feathering them together in any overlap region.
In doing this,
we are effectively approximating a sphere by a large number of facets.

Note that each facet will have a different Fourier sampling
function from the others, and so the point-spread function will vary
from facet to facet. This
change will be quite large between the planet's pole and equator,
and will also be large between different longitudes if only a fraction
of a planet's rotation is observed. The map stitched from
facets cannot be deconvolved with any conventional technique (these
assume a position-independent point-spread function). Although
deconvolution can be done before the facets are stitched together,
this is not optimal as it cannot deconvolve sidelobes originating from
emission in other facets. Making large, overlapping facets will tend
to mitigate this, and the changing geometry will tend to smear out
distant emission anyway.  It will be more of an issue where the Fourier
sampling in each facet is poor - this is not the case in the data we
present. However a more sophisticated joint algorithm
could be expected to give some advantage. For example, an approach
similar to that used by Cornwell and Perley (1992) is possible. Their
algorithm was one used to deconvolve radio images of the sky when the
curvature of the celestial sphere becomes appreciable. They did this
by approximating the celestial sphere as a collection of facets.  In
the results we present below, we have simply deconvolved facets
individually.

Because the size of the point-spread function varies between facets,
so does the brightness sensitivity. As a map in temperature units is
generally more relevant in planetary work, the flux units of the
facets have been converted to brightness temperature units before the
stitching operation.

An alternative approach to producing a longitude-resolved map of the planet
would be to make images of the entire disk at each sampling instant of time,
to deconvolve each of these images, regrid them to a common mapping geometry
and then to perform a weighted sum. This technique has the
advantage that at each time instant, each image is related to the
true image
by a convolution relationship:
a conventional deconvolution algorithm could be used to remove the effect
of a poor Fourier coverage. Note, however, that this approach
would be computationally significantly more expensive. However, more 
importantly,
this approach usually fails as a result of
two characteristics: that the deconvolution
operation in radio interferometry is by necessity a non-linear operation,
and such deconvolution algorithms perform poorly when the signal-to-noise
in the image is very poor and/or when a complex object is imaged with poor
Fourier sampling. Unfortunately the image formed from a single time
instant will have poor sensitivity and poor Fourier coverage. Our
approach avoids these pitfalls by performing the geometric manipulations
before the deconvolution step: each facet is a result of all available
visibility data.

\bigbreak

{\bf 3. Observations and Data Analysis}

As detailed in dP2001, in order to `mark' the location of the Galileo
probe entry on a radio image, we observed the planet with the VLA as
close to the probe entry time as possible. Unfortunately, when the
probe entered Jupiter's atmosphere on December 7, 1995, the planet was
close to conjunction (Dec. 19, 1995) and the array was in 
the B-configuration, so that much of Jupiter's emission would be significantly
resolved out at
short wavelengths. So this period was less than ideal to image Jupiter's
thermal emission.
Observers at optical and infrared wavelengths
also could not observe the planet with high spatial resolution because of its close proximity to
the Sun. Therefore, IRTF (InfraRed Telescope Facility) images taken by
Orton \et (1998) and Ortiz \et (1998) to determine the probe entry
point in relation to infrared hot spots were taken on and before 21
November 1995 and started again on 22 January 1996.  From the long
(many years) time series of images Orton \et and Ortiz \et determined
the most likely appearance of the probe's entry site, or hot spot
through which it entered on Dec. 7 through an interpolation and
extrapolation of their data. After applying a drift correction rate of
103.5 m/s for the period March -- December 1995 and of 102.5 m/s from
January -- May 1996, features at the latitude of the probe entry site
were visible at the same location (in System III coordinates) in all
their images. The infrared intensity, and hence possible dryness, and
the shape of the hot spot varied considerably, however, on timescales
of weeks.

On January 25, approximately 1.5 months after the probe's descent when the
VLA had partially moved to the more compact C-configuration, we observed
the planet at a wavelength of 2 cm. The resolution (FWHM) in this
configuration is $\sim 1-1.5^{\prime\prime}$, and the array was sensitive
to structures up to $\sim 90^{\prime\prime}$. At this wavelength Jupiter's
synchrotron radiation is very low compared to its thermal emission
($\simlt$ 3\%; Bolton \et 2002; de Pater and Dunn, 2003), and one probes
below the ammonia-ice clouds, which makes observations at this wavelength
ideal for imaging the planet's thermal emission. We observed the planet
for a total of $\sim 6-7$ hours. As mentioned in Section 2, we first
constructed a model of a uniform oblate spheroid which best matched the
data; this model was subtracted from the data before we applied the imaging
technique.  This model is very similar to that determined in dP2001: the
disk-averaged temperature is 150 K (the uncertainty is about 5 K), and
limb darkening parameter $p = 0.05$, where limb darkening is parameterized as: $I_0 {\rm cos}^p
(\theta)$, with $I_0$ the brightness temperature and $\theta$ the angle
between the line of sight and the normal to the surface (Butler and
Bastian, 1999).

As described in Section 2 and the Appendix, we imaged facets on
Jupiter every 10 degrees in latitude and longitude. Each facet was
$60^\circ$ in size on Jupiter.  Each facet was deconvolved using a 
CLEAN algorithm (H\"ogbom 1974)
and the intensity units expressed in brightness temperature, using the
appropriate resolution of that facet to convert from Jy to Kelvin. The facets were
then stitched together: the resulting map is shown
in Fig. 2. Since our observing run did not cover Jupiter's full
rotation, the map does not extend over the full 360\deg in longitude.
Fig. 3 shows a similar map of the point-spread functions; this gives the reader an idea
of the change in point-spread function size as it varies with location on
the map.

\bigbreak

{\bf 4. Discussion}

{\it Comparison with IRTF data}

The resemblance of our radio map (Fig. 2) with infrared
images as presented by Orton \et (1998) and Ortiz \et (1998) is
striking. Series of hot spots show up at jovigraphic latitudes between
$7-11^\circ$, i.e., in the same latitude band as the hot spots seen at
infrared wavelengths. The bright spot at a West longitude of 19.7\deg
(Sys. III), and jovigraphic latitude 9.5\deg N ($=$ jovicentric
latitude of 8.3\deg N) is most likely the hot spot through which the
Galileo probe entered; henceforth referred to as the `Galileo' hot
spot. Orton \et (1998) and Ortiz \et (1998) showed that the spots in
this latitude range typically move at a rate between 99 and 105 m/s in
the prograde direction. If the average wind velocity between Dec. 7
1995 and Jan. 25 1996 was 101.5 m/s, the hot spot through which the
Galileo probe entered would have been exactly at the longitude we see
the brightest spot in our map. Orton \et (1998) show that a drift rate
of 103 m/s does best match the IRTF observations between November 1995
and January 1996. Fig. 4 shows a comparison of the NEB as presented in
Fig. 2 (bottom figure) with IRTF data from Ortiz \et (1998) (top
figure). The IRTF image was taken on 22 January 1996, at a central
meridian longitude $\lambda_{III} = 21^\circ$.  We deprojected the
IRTF image onto a grid similar in size to the grid of the radio
image. At a drift rate of 103 m/s, features in this latitude band
should have moved 20.6\deg between the time the IRTF image was taken
(22 Jan. 1996 at $\sim$ 19:35 UT) and the midpoint of the radio image
(25 Jan. 1996 at $\sim$ 16:30 UT). The features in the two images are
thus lined up with respect to each other, assuming a 103 m/s wind
velocity. As shown, each infrared hot spot within $\sim$ 45\deg from
the center shows up as a radio hot spot. We see a few more radio hot
spots at larger distances from the center, which would put the
infrared counterparts too close to the limb on the infrared image for
proper deprojection.

A comparison of the images in Figure 4 clearly shows that the infrared
hot spots are also hot spots at radio wavelengths. The main source of
opacity at infrared wavelengths are cloud particles: hot spots are
known to be devoid of clouds. At radio wavelengths the main source of
opacity is ammonia gas. Because the brightness temperature in the hot
spots is much larger than in surrounding regions, the ammonia gas abundance
must be less so that deeper warmer layers in Jupiter's atmosphere are
probed. Since ammonia gas is the condensible gas in Jupiter's upper
troposphere, we say, in analogy to Earth's atmosphere, that hot spots on
Jupiter apparently contain `dry air'.
In the following section we will derive the ammonia abundance in
the Galileo hot spot, and compare this with the overall ammonia
abundance in the NEB and EZ.

{\it Ammonia Abundance}

The brightness temperature in the Galileo hot spot is $\sim 170$ K, or
$\sim$ 20 K above the general background, and $\sim$ 25 K hotter than
the Equatorial Zone. Below we will work with differences in brightness
temperature rather than the absolute values, since missing short
spacings in the data tend to decrease the overall brightness
temperature. dP2001 performed calculations to derive the
most plausible longitude-averaged NH$_3$ abundance profile in the NEB
and EZ, based upon observations at 2, 3.6 and 6.2 cm. They concluded
that the NH$_3$ abundance in the NEB is roughly $6 \times 10^{-5}$
down to pressure levels of $\sim$ 6 bar, with an NH$_3$ abundance in
the EZ of $\sim 1 \times 10^{-4}$ at $P<4$ bar; the NH$_3$ increases
at larger depths as measured by the Galileo probe. With a temperature
difference between the hot spot and the EZ of $\sim$ 25 K at 2 cm, the
NH$_3$ abundance in the hot spot must be approximately half that in
the NEB, or $\sim 3 \times 10^{-5}$, down to $P>4$ bar. Since we do
not have longitude-resolved images at longer wavelengths we cannot
determine the depth down to which this low NH$_3$ abundance does
extend.

NFR data from the Galileo probe have been
analyzed by Sromovsky \et (1998), and their derived ammonia profile is
shown in Fig. 5 (dashed line). Since the probe entered the hot spot,
this profile should be appropriate for the hot spot at the time of
probe entry. We further indicated on this figure the hot spot profile
as measured with the Infrared Space Observatory (ISO; Fouchet \et
2000a). The latter profile is also quite similar to that presented by Fouchet \et
(2000b) based upon 5 \ms FTS data taken with the Canada-France-Hawaii
telescope (CFHT). The general figure and the weighting functions at
the various radio wavelengths were taken from dP2001.

We have tried to match the radio data with the NFR ammonia profile,
but have not succeeded. In order to match the disk-averaged brightness
temperature at 2--6 cm wavelength, the global ammonia abundance cannot
be above $\sim 1 \times 10^{-4}$ at $P<2-3$ bar. Since the EZ is not a
particularly cold region at radio wavelengths, the NH$_3$ abundance in
the EZ cannot be over $1 \times 10^{-4}$. This profile is indicated on
Fig. 5. At larger depths we adopted the ammonia profile as measured by
the Galileo probe (Folkner \et 1998).  As already mentioned in dP2001,
the ammonia abundance in the NEB should be $\sim 6 \times 10^{-5}$
down to a depth of 5--6 bars. This profile is also indicated on
Fig. 5.  To match the Galileo hot spot temperature at a wavelength of
2 cm, the ammonia abundance needs to be decreased further to $3 \times
10^{-5}$. Due to a lack of longitude resolved images at 3.6 and 6 cm
we cannot determine the depth down to which this low NH$_3$ abundance
extends; based upon the 2 cm data alone we know that it must extend
down to at least the 4 bar pressure level. This profile is also
indicated on Fig. 5. It does match the ISO profile at pressure levels
$P<2$ bar extremely well, but our derived NH$_3$ abundance at
pressures between 2 and 4 bar is typically lower than indicated by the
ISO and NFR data. This region is best probed at radio wavelengths of
3 -- 6 cm.

\bigbreak

{\bf 4. Conclusions}

We have presented the first longitude-resolved map of Jupiter's thermal
radio emission at a wavelength of 2 cm. This map clearly shows the
presence of radio-bright hot spots. A comparison with IRTF images
shows that each radio hot spot is also a hot spot at infrared
wavelengths. Hence the hot spots must indeed lack clouds (infrared
wavelengths) and the air must be dry (low NH$_3$ abundance).  We
derived an ammonia abundance in the hot spot where the probe went down
of approximately $3 \times 10^{-5}$ down to at least the 4 bar level;
we need measurements at longer wavelengths to determine how deep it
extends. This abundance is about half the value derived for the
NEB, and three times less than that in the EZ, as derived from
longitude-averaged images at 2, 3.6 and 6 cm wavelength by dP2001. Our
findings agree at pressures $P \simlt 2$ bar with the infrared
observations of Fouchet \et (2000a, b). However, at pressures between
2 and 4 bar our derived NH$_3$ profile in the hot spot is less than
the values suggested by Fouchet \et from infrared observations, and
also less than the NH$_3$ abundance derived by Sromovsky \et (1998)
from Galileo's NFR experiment.

It is desirable to obtain longitude-resolved images of Jupiter at
wavelengths of 3.6 and 6 cm. Unfortunately, the 3.6 and 6 cm data
presented by dP2001 were snapshots taken on different
days. Differential rotation will tend to smear out features if
differential rotation is not taken into account. A more difficult
problem, however, is Jupiter's synchrotron radiation, which, relative
to the thermal radiation, increases dramatically towards the longer
wavelengths.  We have therefore not attempted to apply this
technique to longer wavelength data.

\bigbreak

{\bf Appendix A: Geometry}

To image a facet, we would like a set of visibilities that views the
facet face on. Instead the visibility data generally corresponds to
an oblique view.  Our aim then is to correct the visibilities for this
distorted view before imaging.  In this way, each visibility can
potentially contribute to a facet.  This appendix addresses two
aspects: the distortion produced by the viewing geometry, and how we
can account for this distortion in the visibility data before imaging.

As an example of the effect of viewing geometry, consider a
cross lying on the surface of a sphere.
Figure 1a shows four identical crosses on a sphere, whose appearance is
different because of the viewing geometry. Figure 1b shows the same four
crosses side by side to accentuate the differences. 
Provided they are sufficiently small, the difference between the cross
viewed face on and the other three crosses
can be approximated as a linear distortion and translation in
the coordinate system. The distorted coordinates of the cross, ${\bf x}'$ are
related to the ``face-on'' coordinates, ${\bf x}$ as

$$
\left(\matrix{x'\cr y'\cr}\right) = 
{\bf D} \left(\matrix{x\cr y\cr}\right) + \left(\matrix{x_0\cr y_0\cr}\right),
\eqno(1)$$

\noindent were ${\bf D}$ is a $2\times 2$ matrix.
To determine ${\bf D}$,
we compute the appropriate transformation in three-dimensional space that
represents the change in the viewing geometry, and then
project this onto our two-dimensional view. 

Consider a facet centered on the origin
in the x-y plane, which corresponds to a face-on view of a region
around the point on Jupiter's surface at longitude, latitude and radius 
of $(\lambda_J,\phi_J,r)$.
Assume the sub-Earth point at the time of interest is $(\lambda_E,\phi_E)$.
We use a Cartesian coordinate system with the $x$-axis lying in
the equatorial plane, the $y$-axis pointing to the pole, and the $z$-axis
pointing towards the observer.
Six transformations are needed to take a facet centered at the origin
to the place on Jupiter as dictated by our viewing geometry. These are:

1. Distort by a shearing operation, ${\bf S}$, to account for 
differential rotation:

$$
S=\left(\matrix{1&s&0\cr 0&1&0\cr 0&0&1\cr}\right).
\eqno(2)$$

\noindent The skew parameter, $s$ will depend on the differential rotation rate at
different latitudes and the elapsed time from the time of interest to
some reference time.

2. Rotate the facet around the $x$-axis by an angle, $\Delta\phi_J$,
which is equal to the difference
of the jovigraphic and jovicentric latitude of the facet center.
This tilting operation
causes the facet to lie tangential to the surface of the planet after
subsequent transformations. We label this rotation ${\bf R_1} = {\bf R}_x(\Delta\phi_J)$.

3. Shift the center of the facet in the direction of the observer so as to 
make the origin correspond to the center of the planet.

$$
{\bf r} = \left(\matrix{0\cr 0\cr r\cr}\right).
\eqno(3)$$

\noindent Note, as Jupiter is oblate, $r$ is a function of the latitude of the
facet center.

4. Rotate about the $x$-axis by the
the jovicentric latitude, $\phi_J$ (i.e. rotate the feature to the correct
latitude). Represent this by ${\bf R_2} = {\bf R}_x(\phi_J)$.

5. Rotate about the $y$-axis by the difference between the longitude of
interest and the sub-Earth longitude. This rotates the feature to the
correct longitudinal view. Represent this by ${\bf R_3} = {\bf R}_y(\lambda_J-\lambda_E)$.

6. Rotate about the $x$-axis by an angle being the negative of the
(jovicentric) sub-Earth latitude. Represent this by ${\bf R_4} = {\bf R}_x(-\phi_E)$

The overall transformation is

$$
{\bf x}' = {\bf R_4}\,{\bf R_3}\,{\bf R_2}\,{\bf R_1}\,{\bf S}\,{\bf x} + 
	   {\bf R_4}\,{\bf R_3}\,{\bf R_2}\,{\bf r}
\eqno(4)$$

Ultimately we are interested only in the projection on the
plane of the sky, as indicated in eq.~1. The distortion matrix of interest
to us, ${\bf D}$
is the upper-left $2\times2$ sub-matrix of ${\bf R_4}\,{\bf R_3}\,{\bf R_2}\,{\bf R_1}\,{\bf S}$. Similarly, our offset 
$(x_0,y_0)^T$ is the $x$ and $y$ component of ${\bf R_4}\,{\bf R_3}\,{\bf R_2}\,{\bf r}$.

Given the image plane transformation that models the viewing geometry,
we can determine the effect on the visibility by using the following
Fourier theorem:

{\narrower
For a Fourier pair,
$$
f({\bf x}) \Leftrightarrow F({\bf u}), \eqno(5)
$$
and a linear transformation of the coordinate system
$$
{\bf x}' = {\bf D}{\bf x},\eqno(6)
$$
then the following is a Fourier transform pair
$$
f({\bf D^{-1}x'}) \Leftrightarrow \det({\bf D})F({\bf D^{T}u'}). \eqno(7)
$$
where ${\bf u}'$ is the Fourier transform coordinates corresponding to
${\bf x}'$.
}

In our situation, we
measure visibilities in the distorted frame, i.e. what we measure
is $\det({\bf D})F({\bf D^{T}u'})$.
Consequently,
correcting the viewing distortion in the visibilities consists of
producing `corrected' Fourier coordinates,
$$
\left(\matrix{u\cr v\cr}\right) = {\bf D^T}\left(\matrix{u'\cr v'\cr}\right), \eqno(8)
$$
and dividing the correlations (and the recorded rms noise) by $\det(D)$.
The correlations also have to be phase rotated, to account for the
shift of $(x_0,y_0)^T$.

The determinant, $\det(D)$, is related to the change in projected area
of the facet when viewed obliquely.  For a feature approaching the limb
of the planet, the projected area approaches zero, and hence dividing by
the determinant amplifies the noise level.  When a feature is behind the
planet, the projected area becomes negative.  We have arbitrarily
discarded visibilities when the determinant drops below 0.1 (i.e. an incident
angle of $84^\circ$. 
Additionally, in the imaging step, we have weighted visibility data by the
reciprocal of the noise variance.  In doing this, we weight down
visibility data corresponding to when a feature is near the limb of the
planet. 

\bigbreak

{\it Acknowledgements}

We thank Glenn Orton and Jose Ortiz for providing us with the IRTF
image, a portion of which is shown in Fig. 4.  This research has in
part been funded by NASA grant NAG5-12062 to the University of
California in Berkeley.

\vfill
\eject

{\bf References}

\pp Anders, E., and N. Grevesse, 1989, Abundances of the elements: meteoritic 
and solar, {\it Geochimica et Cosmochimica Acta} Vol. {\bf 53}, 197-214

\pp Atreya, S.K., T. Owen, and M. Wong, 1996.
Condensible Volatiles, Clouds, and Implications for Meteorology in the Galileo Probe Entry Region: Jupiter Is Not Dry!. {\it BAAS} {\bf 28}, 1133.

\pp Bolton, S. J. and 20 additional authors, 2002. Ultra-relativistic
electrons in Jupiter's radiation belts. {\it Nature} {\bf 415},
987-991

\pp Briggs, D., 1995, High fidelity deconvolution of moderately
resolved sources, PhD Thesis, New Mexico Institute if Mining and 
Technology, Socorro NM (http://www.aoc.nrao.edu/ftp/dissertations/dbtriggs/diss.html)

\pp Butler, B.J. and T.S. Bastian, 1999. Solar system objects. 
{\it Synthesis Imaging in Radio Astronomy II}, {\it ASP Conference
series} {\bf 180}. Eds. G.B. Taylor, C.L. Carilli, and R.A. Perley.

\pp Cornwell, T.J., and R.A. Perley, 1992. Radio-interferometric imaging of
very large fields -- The problem of non-coplanar arrays. {\it
Astron. Astrophys.} {\bf 261}, 353-364.

\pp de Pater, I., 1980. 21 cm maps of Jupiter's radiation belts from all 
  rotational aspects, {\it Astron. Astrophys.} {\bf 88}, 175-183.

\pp  de Pater, I. and D.E. Dunn, 2003.  VLA Observations of Jupiter's
Synchrotron Radiation at 15 and 22 GHz. {\it Icarus} {\bf 163}, 449-455.

\pp de Pater, I., D. Dunn, K. Zahnle and P.N. Romani,
2001. Comparison of Galileo Probe Data with Ground-based Radio
Measurements. {\it Icarus}, {\bf 149}, 66-78 (dP2001)

\pp Folkner, W.M., R. Woo, and S. Nandi, 1998. Ammonia abundance in
Jupiter's atmosphere derived from the attenuation of the Galileo
probe's radio signal. {\it J. Geophys. Res. Planets}, {\bf 103}, 22847-22856.

\pp Fouchet, T., E. Lellouch, B. Bezard, T. Encrenaz, and P. Drossart,
2000.  ISO-SWS observations of Jupiter: measurement of the ammonia
tropospheric profile and of the $^{15}$N/$^{14}$N isotopic ratio. {\it
Icarus} {\bf 143}, 223-243.

\pp Fouchet, T., E. Lellouch, J.-P. Maillard, B. Bezard, C. Cottaz, and I. Kleiner,
2000. Determination of Jupiter's N/H ratio from FTS observations at 5
micron. {\it BAAS} {\bf 32}, \# 12.16

\pp Friedson, A.J. and G.S. Orton, 1999. A dynamical model of Jupiter's
5-micron hot spot. {\it BAAS} {\bf 31}, 1155.

\pp Grevesse, N., D.L. Lambert, A.J. Sauval, E.F. van Dishoeck,
C.B. Farmer, and R.H. Norton, 1991, Vibration rotation bands of CH in
the solar infrared spectrum and the solar carbon abundance,
Astron. Astrohys.  {\bf 242}, 482-495.

\pp H\"ogbom, J.A., 1974, Aperture synthesis with a non-regular distribution
of interferometer baselines, {\it Astron. Astrophys}, {\bf 15}, 417-426.
\pp Mahaffy, P.R., H.B. Niemann, and J.E. Demick, 1999.
Deep Atmosphere Ammonia Mixing Ratio at Jupiter from the Galileo Probe
Mass Spectrometer.  {\it BAAS} {\bf 31}, 5205.

\pp Ortiz, J.L., G.S. Orton, A.J. Friedson, S.T. Stewart, B.M. Fisher, and 
J.R. Spencer, 1998, Evolution and persistence of 5-$\mu$m hot spots at
the Galileo probe entry latitude. {\it J. Geophys. Res.} {\bf 103},
23,051-23,069.

\pp G.S. Orton, B.M. Fisher, K.H. Baines, S.T. Stewart, A.J.
Friedson, J.L. Ortiz, M. Marinova, M. Ressler, A. Dayal, W. Hoffmann,
J. Hora, S. Hinkley, V. Krishnan, M. Masanovic, J. Tesic, A.  Tziolas,
and K.C. Parija, 1998.  Characteristics of the Galileo Probe entry
site from Earth-based remote sensing observations. {\it
J. Geophys. Res.} {\bf 103}, 22791--22814.

\pp Sault, R.J., T. Oosterloo, G.A. Dulk, and Y. Leblanc, 1997, The
first three-dimensional reconstruction of a celestial object at radio
wavelengths: Jupiter's radiation belts, {\it Astron. Astrophys}, {\bf
324}, 1190-1196.

\pp Showman, A.P. and T.E. Dowling 2000. Nonlinear simulations of
Jupiter's 5-micron hot spots.  {\it Science} {\bf 289,} 1737-1740.

\pp Showman, A. P. and A.P. Ingersoll, 1998.
Interpretation of Galileo Probe Data and Implications for Jupiter's Dry Downdrafts.
{\it Icarus} {\bf 132}, 205-220.
 
\pp Sromovsky, L.A., A.D. Collard, P.M. Fry, G.S.  Orton, M.T. Lemmon,
M.G. Tomasko, and R.S.  Freedman, 1999.  Galileo Probe measurements of
thermal and solar radiation fluxes in the jovian atmosphere.  {\it
J. Geophys. Res.} {\bf 103}, 22929--22978.

\pp Wong, M.H., S.K. Atreya, and P.N. Romani, 1998.
Entrainment in the Galileo Probe Site Downdraft. {\it BAAS}, {\bf 30}, 1075.

\vfill\eject
\centerline{\epsfbox{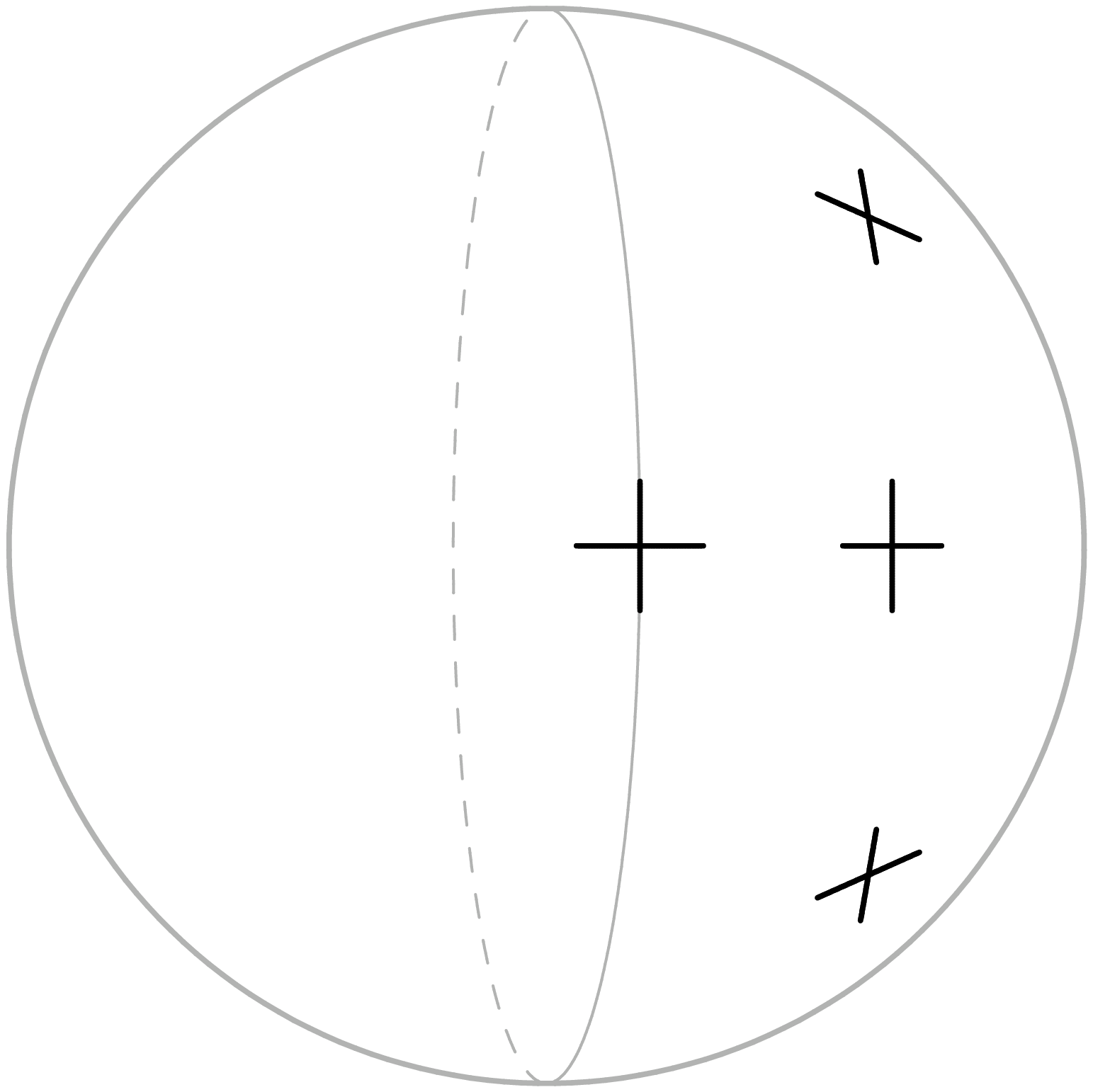}}

\bigskip

\centerline{\epsfysize=1.3cm\epsfbox{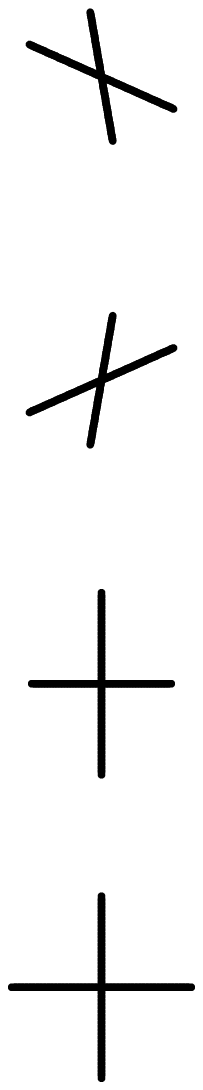}}

Fig. 1. Four identical crosses on the face of a planet. The viewing
geometry distorts the appearance of the crosses -- Fig 1b presents the
crosses side-by-side to help accentuate the distortions.

\vfill\eject
\centerline{\epsfxsize=14cm\epsfbox{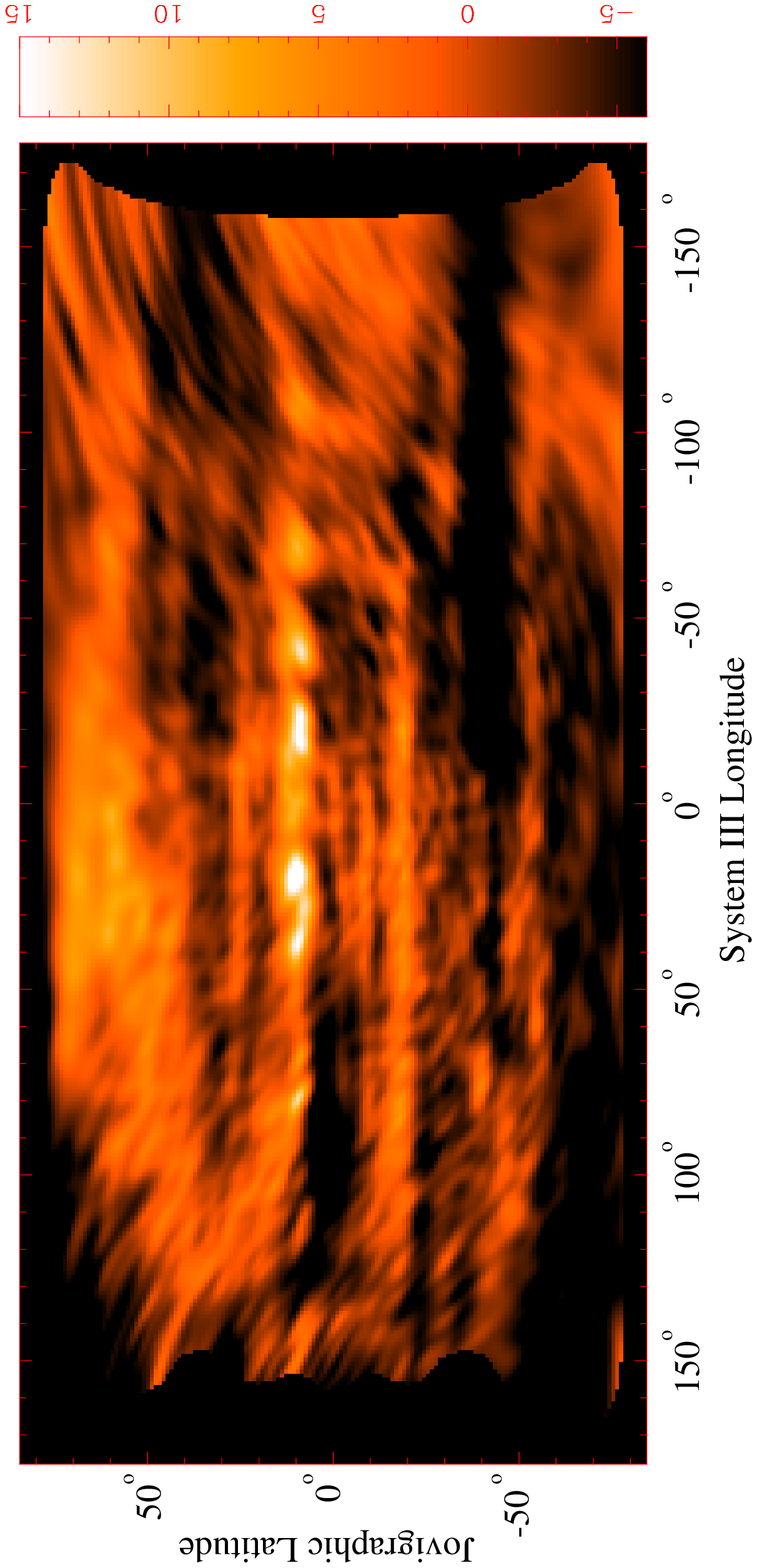}}

Fig. 2. Radio map of Jupiter at a wavelength of 2 cm. The
data were obtained with the VLA on 25 January 1996, and processed as
described in this paper. Note that a uniform disk was subtracted, so
only deviations in temperature from a 150 K disk are shown.

\vfill\eject
\centerline{\epsfxsize=14cm\epsfbox{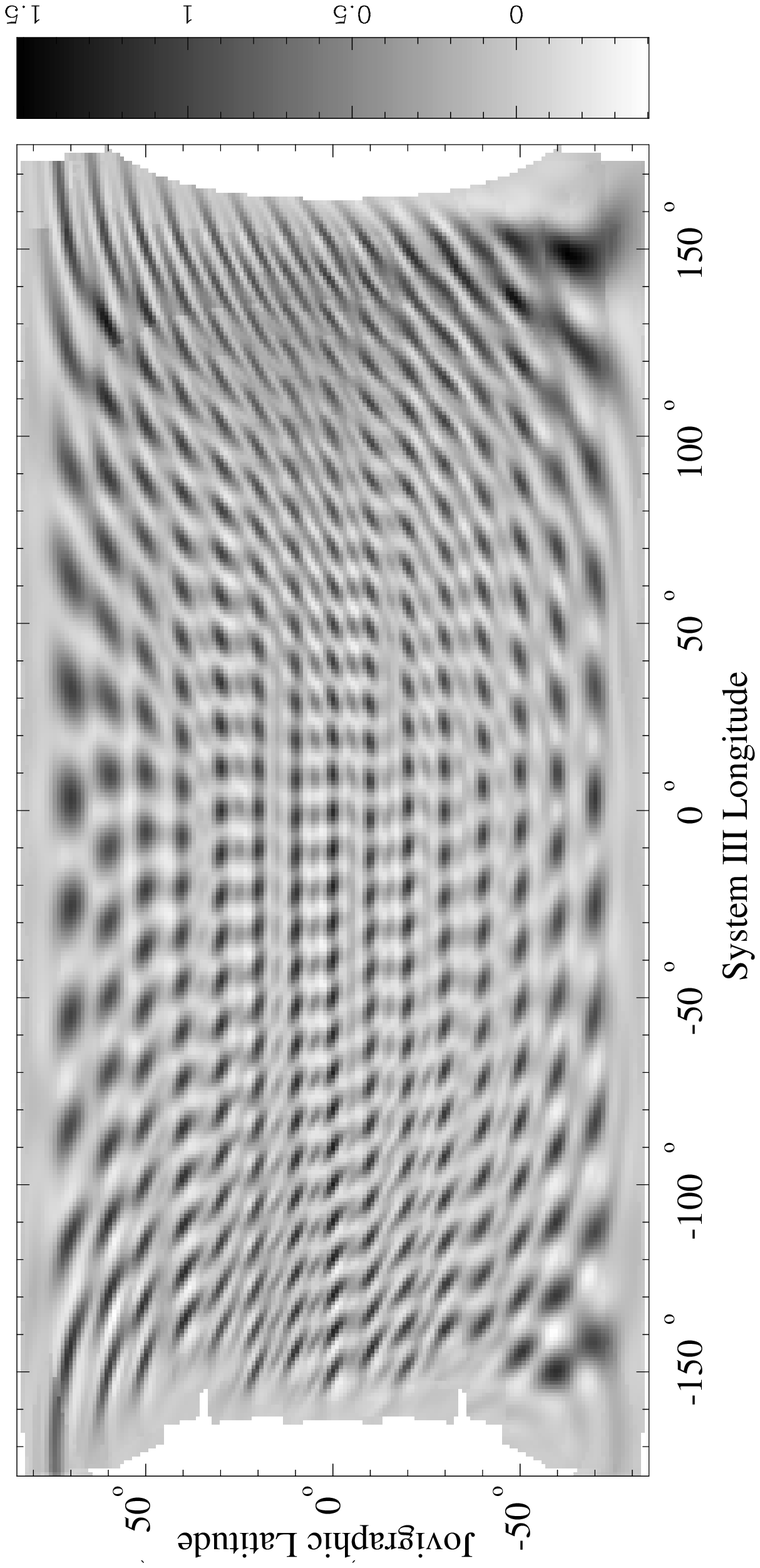}}

Fig. 3. A representation of the point-spread functions for each facet
imaged on the planet.

\vfill\eject
\centerline{\epsfxsize=14cm\epsfbox{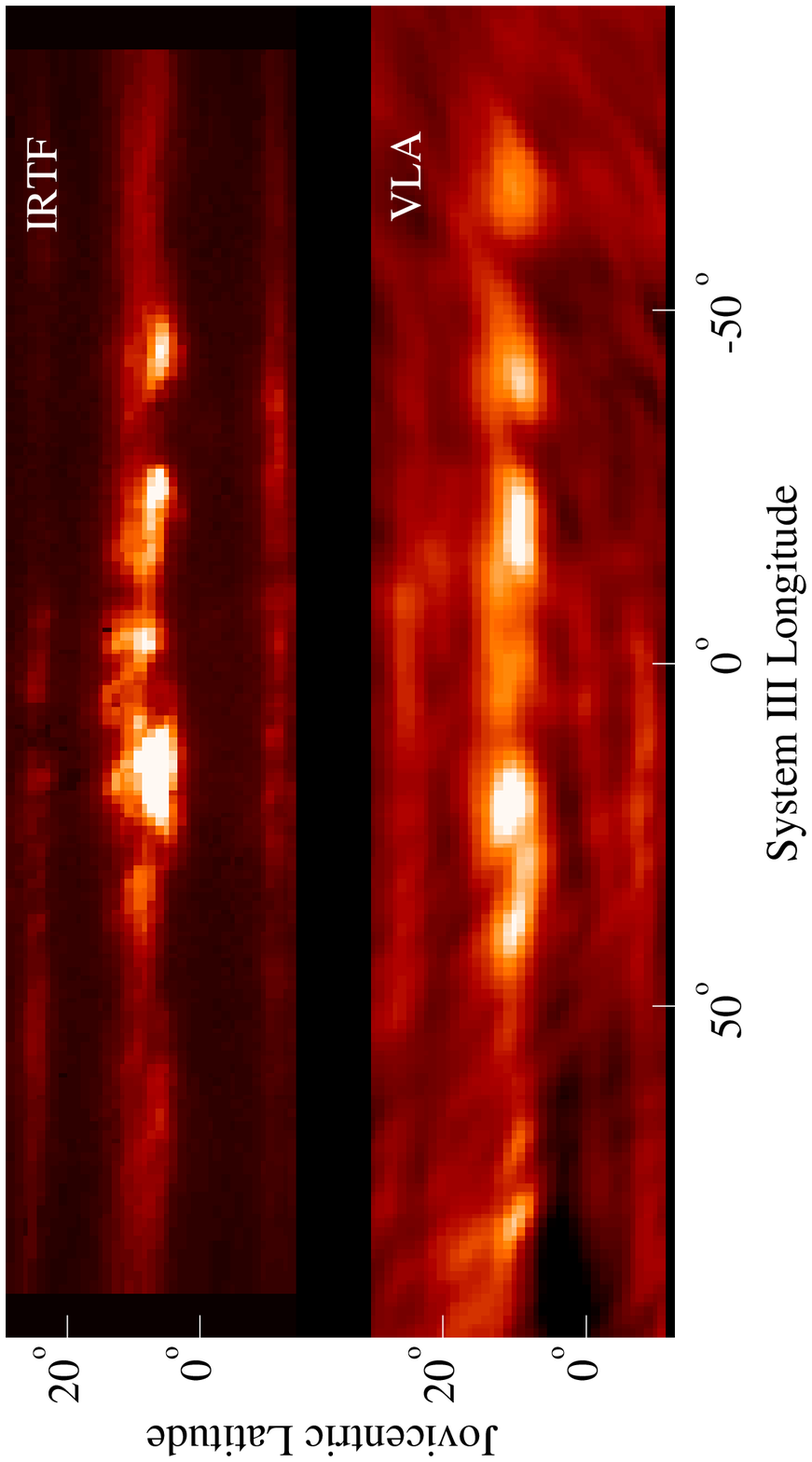}}

Fig. 4. Comparison of an IRTF image (top) with our radio image
(bottom). We show a $\sim$ 45\deg latitude range around the NEB on
both images (indicated in approximate jovicentric coordinates), and a
180\deg view in longitude (in System III coordinates).  The IRTF image
was taken on 22 January 1996 by J.L. Ortiz and G.S. Orton (see Ortiz
\et 1998 and Orton \et 1998). The central meridian longitude in the
IRTF image $\lambda_{III}=21^\circ$. With a wind velocity of 103 m/s,
the features in the two images line up well.

\vfill\eject
\centerline{\epsfxsize=12cm\epsfbox{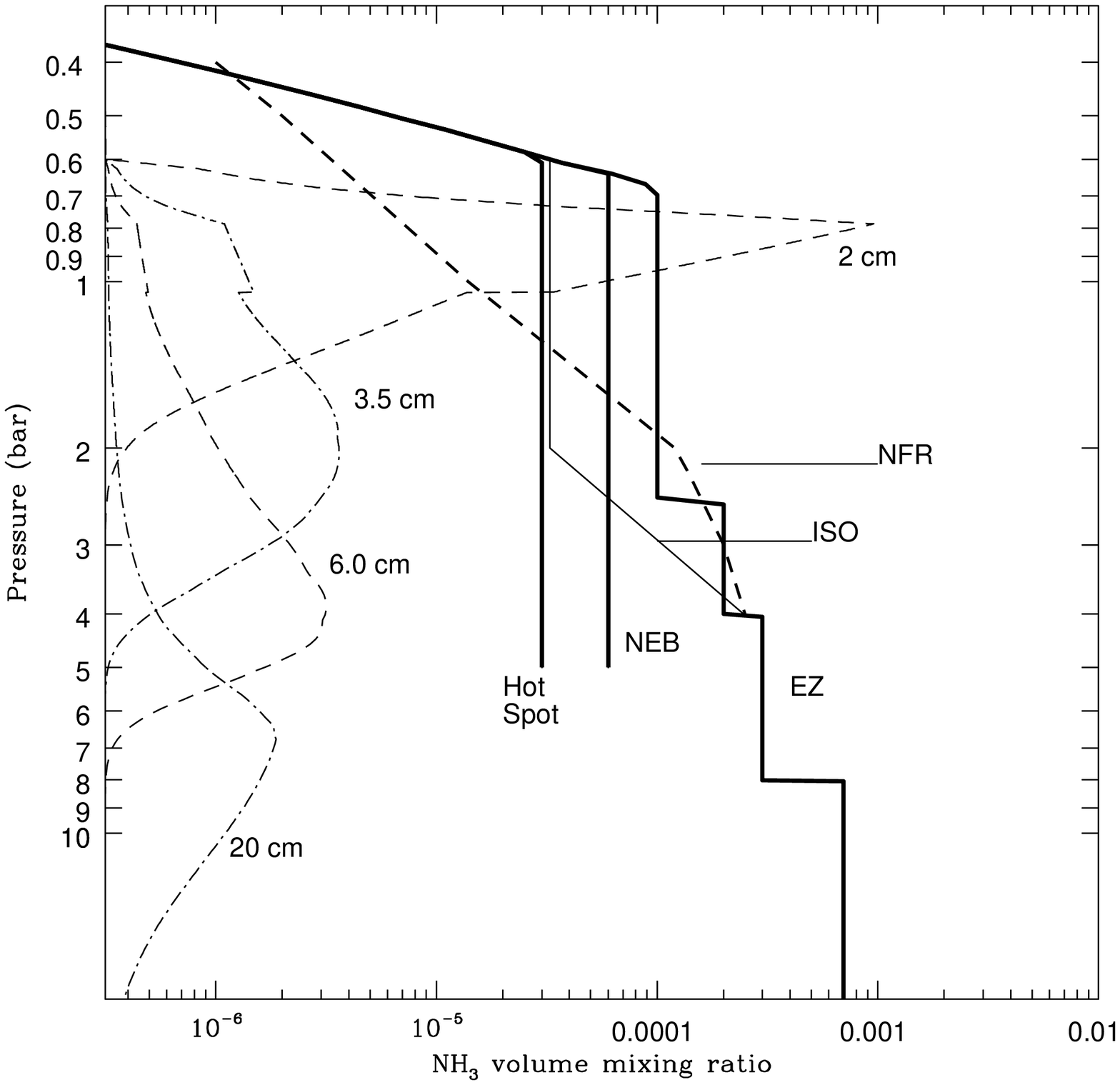}}

Fig. 5. Graph of the ammonia abundance profiles as derived from the
radio data in the EZ, NEB, and Galileo hot spot (see text for detailed
explanations). The profile as derived from the NFR data is indicated
by the dashed line, and that obtained with ISO by the thin solid
line. Weighting functions for the different wavelengths are indicated
on the left of the figure (adapted from dP2001).

\end